\documentclass[11pt,a4paper]{article}

\pdfoutput=1
\usepackage{jcappub}
\usepackage{graphicx}% Include figure files
\usepackage{dcolumn}% Align table columns on decimal point
\usepackage{bm}% bold math
\usepackage{color}
\usepackage{calrsfs}
\usepackage{hyperref}
\usepackage{multirow}
\usepackage{float}
\input epsf

\newcommand{\be}{\begin{equation}}
\newcommand{\e}{\end{equation}}
\newcommand{\bear}{\begin{eqnarray}}
\newcommand{\ear}{\end{eqnarray}}

\begin{document}

\title{Cross-correlating 21cm intensity maps with Lyman Break Galaxies in the post-reionization era}

\author[a,b]{Francisco Villaescusa-Navarro,} 
\author[a,b]{Matteo Viel,} 
\author[c]{David Alonso,}
\author[d]{Kanan K. Datta,} 
\author[e]{Philip Bull,} 
\author[f,g,h]{M\'ario G. Santos}

\affiliation[a]{INAF - Osservatorio Astronomico di Trieste, Via Tiepolo 11, 34143, Trieste, Italy}
\affiliation[b]{INFN sez. Trieste, Via Valerio 2, 34127 Trieste, Italy}
\affiliation[c]{Astrophysics Department, University of Oxford, DWB, Keble Road, Oxford OX1 3RH, UK}
\affiliation[d]{Department of Physics, Presidency University, 86/1 College Street, Kolkata -700073, India}
\affiliation[e]{Institute of Theoretical Astrophysics, University of Oslo, P.O. Box 1029 Blindern, N-0315 Oslo, Norway}
\affiliation[f]{Department of Physics, University of Western Cape, Cape Town 7535, South Africa}
\affiliation[g]{SKA SA, 3rd Floor, The Park, Park Road, Pinelands, 7405, South Africa}
\affiliation[h]{CENTRA, Instituto Superior T\'ecnico, Universidade de Lisboa, Lisboa 1049-001, Portugal}

\emailAdd{villaescusa@oats.inaf.it}
\emailAdd{viel@oats.inaf.it}
\emailAdd{david.alonso@astro.ox.ac.uk}
\emailAdd{kanan@ncra.tifr.res.in}
\emailAdd{p.j.bull@astro.uio.no}
\emailAdd{mgrsantos@uwc.ac.za}

\abstract{We investigate the cross-correlation between the spatial distribution of Lyman Break Galaxies (LBGs) and the 21cm intensity mapping signal at $z\sim[3-5]$. At these redshifts, galactic feedback is supposed to only marginally affect the matter power spectrum, and the neutral hydrogen distribution is independently constrained by quasar spectra. Using a high resolution N-body simulation, populated with neutral hydrogen a posteriori, we forecast for the expected LBG-21cm cross-spectrum and its error for a 21cm field observed by the Square Kilometre Array (SKA1-LOW and SKA1-MID), combined with a spectroscopic LBG survey with the same volume. The cross power can be detected with a signal-to-noise ratio (SNR) up to $\sim\!10$ times higher (and down to $\sim\! 4$ times smaller scales) than the 21cm auto-spectrum for this set-up, with the SNR depending only very weakly on redshift and the LBG population. We also show that while both the 21cm auto- and LBG-21cm cross-spectra can be reliably recovered after the cleaning of smooth-spectrum foreground contamination, only the cross-power is robust to problematic non-smooth foregrounds like polarized synchrotron emission.}

\maketitle

\section{Introduction}
\label{sec:introduction}

Radio telescopes can detect the presence of neutral hydrogen (HI) through spectroscopic observations of its redshifted 21cm line. This technique is one of the most promising ways of studying the Epoch of Reionization (EoR), the era in which the first galaxies and stars formed and ionized the Universe. Neutral hydrogen is also abundant after the end of reionization, with the majority of the HI (in mass) expected to be localized within dense galactic or protogalactic environments, and a much smaller amount filling most of the volume and giving rise to a low density photo-ionized cosmic web.

At intermediate redshifts, the spatial distribution of HI can be used to place tight constraints on cosmological parameters \cite{Bharadwaj_2001A, Battye:2004re, Chang_2008,Loeb_Wyithe_2008, Wyithe_2008, Camera_2013, Bull_2014}, with experiments taking advantage of the significantly larger survey volumes that can be accessed compared with low redshift galaxy surveys. The key observational quantity is the 21cm power spectrum, which can be measured directly. On large scales, the shape of the 21cm power spectrum is expected to be the same as that of the underlying dark matter, up to a bias factor. On small scales, the 21cm power spectrum provides information about the spatial distribution of neutral hydrogen within dark matter halos, and can therefore be used to study galaxy formation and evolution.

In order to extract the maximum amount of information from forthcoming observations, it will be necessary to accurately model the HI spatial distribution. The amplitude of the 21cm power spectrum can change significantly depending on the model used \citep{Villaescusa_2014a}, for example, which has important implications for forthcoming radio surveys designed to detect it. These include LOFAR\footnote{http://www.lofar.org/}, the Murchison Wide-field Array\footnote{http://www.mwatelescope.org/}, GMRT\footnote{http://gmrt.ncra.tifr.res.in/}, the Ooty Radio Telescope \cite{Ali_2013}, 
CHIME\footnote{http://chime.phas.ubc.ca/}, ASKAP\footnote{http://www.atnf.csiro.au/projects/askap/index.html}, MeerKAT\footnote{http://www.ska.ac.za/meerkat/} and the Square Kilometre Array (SKA)\footnote{https://www.skatelescope.org/}, which are expected to detect the 21cm power spectrum before, during and after the EoR.

Unfortunately, the radio signal arising from cosmic neutral hydrogen is several orders of magnitude weaker than the emission from galactic and extragalactic foregrounds like galactic synchrotron. These are believed to have smooth frequency spectra, making it possible to model their contribution to the overall signal and subtract them, although this process will always leave some residual contamination. Since most of the foreground signal is expected to be uncorrelated with the cosmological HI, the cross-correlation between the 21cm signal and the spatial distribution of Lyman Break Galaxies (LBGs) at intermediate/high redshifts should be only weakly affected by residual foregrounds, suggesting the cross-spectrum as a robust way of confirming the detection of the 21cm signal. As we shall soon see, the errors on the cross-spectrum are also significantly lower than those of the 21cm auto-power spectrum, making it possible to detect the 21cm signal more easily, and over a wider range of scales, than for the auto-spectrum alone. Indeed, a recent detection of the cosmological 21cm signal at $z\sim1$ was made by cross-correlating intensity maps and optically selected galaxies \cite{Chang_2010,Masui_2013}, and several works have also explored the possibility of using the cross-power spectrum to study the EoR \cite{Furlanetto_2007, Lidz_2009, Wiersma_2013, Park_2013, Mao_2014}.

The aim of this paper is to investigate the LBG-21cm cross-correlation at redshifts $z\sim3-5$, after the end of reionization. In particular, we want to study how the shape and amplitude of the cross-power spectrum depend on redshift, LBG population and the method used to model the distribution of HI, and to predict on which scales the cross-spectrum can be detected by a future 21cm survey on the SKA1-LOW and SKA1-MID radio telescopes. We also investigate how well the cosmological signal can be recovered once the process of foreground cleaning has been carried out, both for the 21cm auto-spectrum and the LBG-21cm cross-spectrum.

In what follows, we model the HI distribution using two different methods that are capable of reproducing the measured column density function of the Damped Lyman Absorber systems (DLAs). One method predicts a 21cm power spectrum that is in excellent agreement with the one obtained by using the pseudo-radiative transfer method of Dav\'e et al. \cite{Dave_2013}, but fails to reproduce recently-measured values of the DLA bias \cite{Font_2012}. The other method is capable of reproducing both the bias and column density distribution function of the DLAs (see \cite{Villaescusa_2014a} for details). The spatial distribution of the galaxy population we consider in this paper, Lyman Break Galaxies (LBGs), is obtained by populating the dark matter halos of an N-body simulation using a simple halo occupation distribution (HOD) model. We then cross-correlate both fields and investigate the detectability of the signal with a future single-field interferometric survey on SKA1-LOW and SKA1-MID.

The paper is organized as follows. In Sec. \ref{sec:21cm} we describe the N-body simulation carried out and the method used to simulate the spatial distribution of HI. The halo occupation distribution (HOD) model employed to simulate the abundance and spatial distribution of the LBGs is explained in Sec. \ref{sec:HOD}, and the LBG-21cm cross-power spectrum, together with predicted observational errors, is presented in Sec. \ref{sec:results}, along with a discussion of the dependence on redshift and galaxy population. We investigate the dependence of our results on the HI distribution model in Sec. \ref{sec:HI_modeling}, and the impact of foreground subtraction on both the 21cm auto-spectrum and the LBG-21cm cross-spectrum in Sec. \ref{sec:fg}. We conclude in Sec. \ref{sec:conclusions}.

\section{Modeling the 21cm intensity field}
\label{sec:21cm}

While hydrodynamic N-body simulations are better suited for 21cm studies, the large box sizes needed to simulate a representative sample of LBGs, together with the high resolution required to resolve the smallest halos that host HI, make it computationally prohibitive to run them. Instead, we will rely on relatively large box size, high-resolution, pure cold dark matter (CDM) simulations.

We have run a pure-CDM N-body simulation, using the TreePM code {\sc GADGET-III} \cite{Springel_2005}. The simulation box size is 120 $h^{-1}$Mpc and the value of the gravitational softening is set to $1/40$ of the mean inter-particle linear spacing; the mass resolution of the simulation is equal to $m_{\rm p}=1.42\times10^{8}~h^{-1}{\rm M}_\odot$, i.e. we have $1024^3$ CDM particles. The input cosmological parameters take the current Planck best-fit values \cite{Planck_2013}: $\Omega_{\rm m}=\Omega_{\rm cdm}+\Omega_{\rm b}=0.3175$, $\Omega_{\rm b}=0.049$, $\Omega_\Lambda=0.6825$, $h=0.6711$, $n_s=0.9624$ and $\sigma_8=0.834$. Initial conditions were set up at $z=99$ by displacing the particle positions from a regular cubic grid using the Zel'dovich approximation, and the simulation was run until $z=3$, with snapshots saved at redshifts 7, 6, 5, 4.5, 4, 3.5 and 3. Dark matter halos were identified using the Friends-of-Friends (FoF) algorithm \cite{FoF} with a linking length parameter of $b=0.2$.

The distribution of neutral hydrogen (HI) is simulated by assigning HI to the particles belonging to halos according to Model 3 of Bagla et al. \cite{Bagla_2010}. In this model, the total HI mass hosted by a dark matter halo only depends on its total mass, and is given by
\begin{equation}  
M_{\rm HI}(M) = \left\{ 
  \begin{array}{l l}	
    f_3\frac{M}{1+\left(\frac{M}{M_{\rm max}}\right)} & \quad \text{$M_{\rm min}\leqslant M$}
    \\
    0 & \quad \text{otherwise},\\
  \end{array} \right.
\label{M_HI_Bagla3}
\end{equation}
where $f_3$, $M_{\rm min}$ and $M_{\rm max}$ are free parameters. The values of $M_{\rm min}$ and $M_{\rm max}$ are obtained, respectively, by assuming that halos with circular velocities smaller than 30 km/s do not host HI; and that all those with circular velocities larger than 200 km/s contain essentially an equal amount of HI. The value of the parameter $f_3$ is set by demanding that $\Omega_{\rm HI}(z)=\rho_{\rm HI}(z)/\rho_{\rm 0, crit}=10^{-3}$, which is the value obtained from the abundance of DLAs and Lyman limit systems (LLS), almost independently of redshift \cite{Peroux_2003, Zwaan_2005, Rao_2006, Lah_2007, Martin_2010, Braun_2012, Noterdaeme_2012, Zafar_2013}. In Ref.~ \cite{Villaescusa_2014a}
 this method was compared against the pseudo-radiative transfer code presented in \cite{Dave_2013}, finding excellent agreement between the two in terms of the amplitude and shape of the 21cm power spectrum. We use this method for modeling the spatial distribution of HI throughout, although some results using a different method are also presented in Section \ref{sec:HI_modeling}.

After obtaining the distribution of neutral hydrogen in comoving real-space as described above, we obtain the redshift-space distribution by displacing the particle positions by $(1+z)\vec{v}_{\parallel}(\vec{r})/H(z)$ along a given axis, where $\vec{v}_{\parallel}(\vec{r})$ is the particle physical velocity along the axis. We then compute the redshift-space brightness temperature fluctuation using \cite{Mao_2012}
\be
\delta T_b(\vec{s},\nu)=\overline{\delta T_b}(z)\left(\frac{\rho_{\rm HI}(\vec{s})}{\bar{\rho}_{\rm HI}}\right)
\left[ 1-\frac{T_\gamma(z)}{T_s(\vec{s})}\right],
\e
where $\rho_{\rm HI}(\vec{s})$ is the density of neutral hydrogen in the redshift-space position $\vec{s}$, $\bar{\rho}_{\rm HI}$ is the mean density of neutral hydrogen, $T_\gamma(z)$ is the CMB temperature at redshift $z$, $T_s$ is the spin temperature characterizing the relative population of HI atoms in different states, and
\be
\overline{\delta T_b}(z)=23.88~\bar{x}_{\rm HI}\left( \frac{\Omega_{\rm b}h^2}{0.02}\right)\sqrt{\frac{0.15}{\Omega_{\rm m}h^2}\frac{(1+z)}{10}}~{\rm mK},
\e
where $\bar{x}_{\rm HI}=\bar{\rho}_{\rm HI}/\bar{\rho}_{\rm H}$ is the average neutral hydrogen fraction. The 21cm power spectrum is defined as $P_{\rm 21cm}(k,z)=\langle \delta T_b(\vec{k},\nu) \delta T^*_b(\vec{k},\nu)\rangle$, and has units of ${\rm mK}^2~(h^{-1}{\rm Mpc})^3$. It is more convenient to work with the normalized 21cm power spectrum,
\be
P^{\rm norm}_{\rm 21cm}(k,z)=\left(\frac{1}{\overline{\delta T_b}(z)}\right)^2 \left\langle \delta T_b(\vec{k},\nu) \delta T_b^*(\vec{k},\nu)\right\rangle~,
\e
which has the same units as the LBG power spectrum and the normalized LBG-21cm cross-spectrum, i.e. $(h^{-1}{\rm Mpc})^3$.

\section{Modeling the LBG distribution}
\label{sec:HOD}

We populate the dark matter halos of our N-body simulation with LBGs, with the purpose of reproducing both their observed number density and their real-space two-point correlation function, as derived from the observational measurements. We focus on the results of Lee et al. (2006) \cite{Lee_2006}, and in particular on the clustering properties of their $U$, $B_{435}$ and $V_{606}$ dropout LBG catalogues. As far as we are aware, it is the only work so far to have studied the clustering properties of the faintest ($z_{850}\leqslant27$) LBGs at redshifts $z=4$ and $z=5$.

The dark matter halos are populated using a simple HOD model with three free parameters: $\alpha$, $M_{\rm min}$ and $M_1$. Halos with masses smaller than $M_{\rm min}$ are assumed to not host LBGs, while the number of LBGs in those with $M\textgreater M_{\rm min}$ is taken to follow a Poisson distribution with a mean given by
\be
\langle N_{\rm LBG} \rangle(M) = \left( \frac{M}{M_1} \right)^\alpha~.
\label{HOD_mean}
\e
The galaxies are then placed on top of the CDM particles belonging to the host dark matter halos, and assigned the same velocities as the particles (i.e. we assume no velocity bias between the LBG distribution and the underlying matter).

For a given galaxy population to simulate, the value of the parameter $M_{\rm min}$ is fixed, once the values of $\alpha$ and $M_1$ are known, by demanding that our LBG mock catalogue reproduces the observed galaxy number density of the chosen population, $n_{\rm LBG}$. Therefore, our HOD model only has two free parameters, $\alpha$ and $M_1$, and their values have to be found requiring that the clustering properties of the selected LBGs reproduce the observations. Once the values of the HOD parameters are selected, we populate the dark matter halos of the simulation with LBGs by employing the method described above, and compute the two-point correlation function (in real-space) using the Landy-Szalay \cite{Landy-Szalay_93} estimator,
\be
\xi(r)=\frac{DD(r)-2DR(r)+RR(r)}{RR(r)}~,
\e
where $DD(r)$ and $RR(r)$ are the normalized number of galaxies and random point pairs in the interval $[r,r+dr]$, respectively, and $DR(r)$ is the normalized number of galaxy-random point pairs. We then quantify the agreement of the clustering properties of the resulting LBG mock catalogue with the observational measurements by calculating
\be
\chi^2=\sum_i \left[\frac{\xi_{\rm mock}(r_i)-\xi_{\rm observations}(r_i)}{\delta \xi_{\rm mock}(r_i)}\right]^2,
\e
where the sum is performed for bins with $r>0.25~h^{-1}$Mpc to avoid contamination from the 1-halo term, and $\delta \xi_{\rm mock}$ is the error on the two-point correlation function. Both observational errors (Table \ref{HOD_table}) and cosmic variance \cite{Mo_1992} contribute to the uncertainty; we select the minimum of the two for each bin in radius. For $\xi_{\rm observations}(r)$ we use the form
\be 
\xi_{\rm observations}(r)=\left(\frac{r}{r_0}\right)^{-\gamma},
\e
with values for $r_0$ and $\gamma$ taken from Table 3 of \cite{Lee_2006} for the particular LBG type selected (see Table \ref{HOD_table}). The parameter values that minimize $\chi^2$ are found by mapping out the $M_1 - \alpha$ plane on an increasingly fine grid until the required accuracy is achieved. For simplicity, we assume that the correlation function datapoints are uncorrelated.\footnote{The resulting best-fit HOD parameters would be unlikely to change much if the full covariance matrix was used; also, we are not interested in the errors on the HOD parameters here.}

\begin{table}
\begin{center}
{\renewcommand{\arraystretch}{1.2} \begin{tabular}{|c|c|c|c|c|}
\hline
\multirow{2}{*}{Redshift} & \multirow{2}{*}{Magnitudes} & $r_0$ & \multirow{2}{*}{$\gamma$} & $n_{\rm LBG}$  \\ 
& & [$h^{-1}$Mpc] & & $[(h^{-1}{\rm Mpc})^{-3}]$ \\ \hline
3 & $R\leqslant25.5$ & $4.0^{+0.2}_{-0.2}$ & $1.63^{+0.11}_{-0.12}$ & $(3.3\pm1.0)\times10^{-3}$\\ \hline
4 & $z_{850}\leqslant27.0$ & $2.8^{+0.2}_{-0.2}$ & $1.69^{+0.16}_{-0.15}$ & $(7.3\pm1.0)\times10^{-3}$\\ \hline
4 & $z_{850}\leqslant26.5$ & $3.7^{+0.3}_{-0.3}$ & $1.70^{+0.17}_{-0.13}$ & $(4.5\pm0.7)\times10^{-3}$\\ \hline
5 & $z_{850}\leqslant27.0$ & $4.2^{+0.4}_{-0.5}$ & $1.85^{+0.20}_{-0.23}$ & $(4.2\pm0.8)\times10^{-3}$\\ \hline
\end{tabular} }
\end{center}
\caption{\label{HOD_table}Clustering properties of the LBG populations considered in this paper (taken from \cite{Lee_2006}).}
\end{table}

\begin{figure}
\begin{center}
\includegraphics[width=0.7\textwidth]{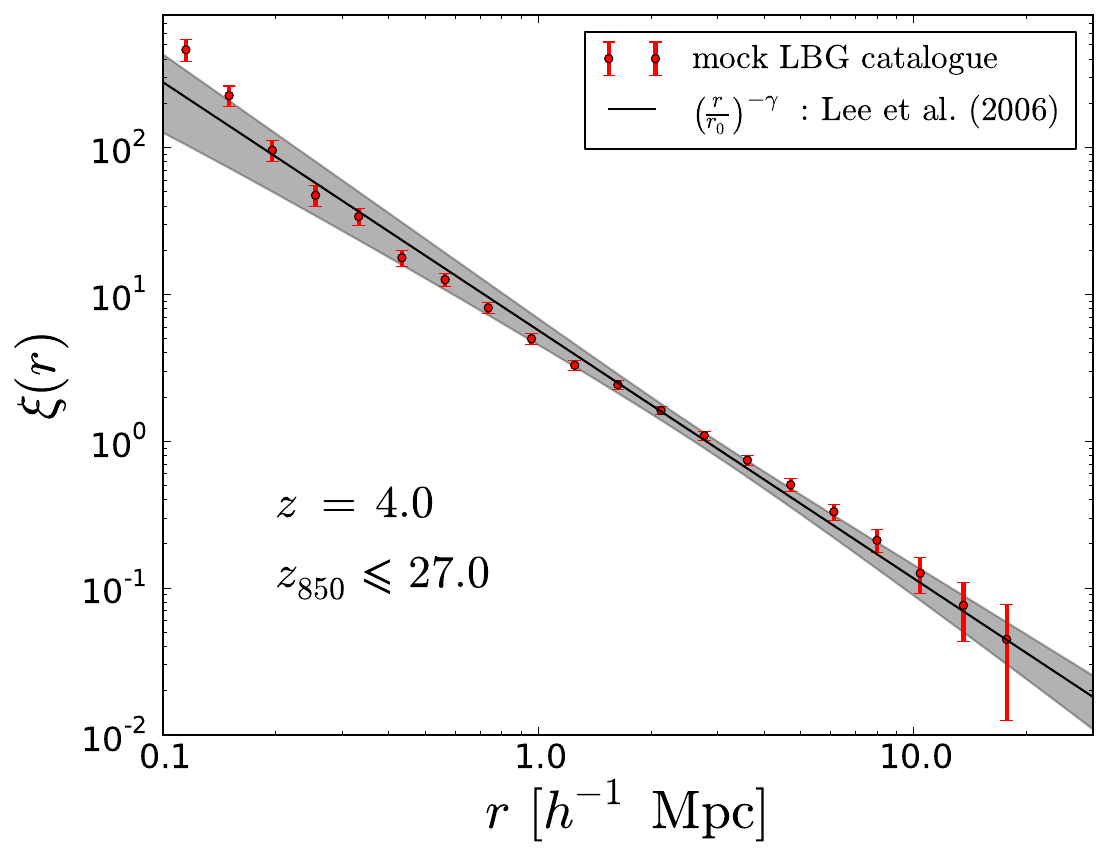}\\
\end{center}
\caption{Real-space correlation function of LBGs with magnitudes $z_{850}\le27$ at $z=4$ ($B_{435}$ dropout) from Lee et al. \cite{Lee_2006} (black line), with corresponding uncertainty (grey shaded region). Red points show the correlation function of our mock LBG catalogue, with errors due to sample variance.}
\label{HOD_CF}
\end{figure}

Following the above procedure, we constructed LBG mock catalogues by populating the dark matter halos of our N-body simulation at $z=3$ (U-dropout) with LBGs with magnitudes $R\le25.5$; at $z=4$ ($B_{435}$ dropout) with magnitudes $z_{850}\le27$ and $z_{850}\le26.5$; and at $z=5$ ($V_{606}$ dropout) with magnitudes $z_{850}\le27$. As an example, Fig. \ref{HOD_CF} shows the real-space correlation function of the $z=4$, $z_{850}\le27$, mock LBG catalogue, together with the inferred real-space correlation function from observations \cite{Lee_2006}. Our mock catalogue reproduces the inferred clustering properties of those galaxies on scales $r>0.25~h^{-1}$Mpc very well. On smaller scales, the clustering properties of the mock galaxies deviates from the large-scale trend due to the clustering of several galaxies within the same halo (the 1-halo term).

\section{LBG-21cm cross-correlation signal}
\label{sec:results}

We now investigate the cross-correlation of the 21cm and the LBG fields. After applying the methods described above to generate the redshift-space distribution of both HI and LBGs, we compute the cross-power spectrum of both fields as
\begin{equation}
P_{\rm LBG-21cm}(k,z)=\left\langle \Re \left( {\delta_{\rm LBG}^*(\vec{k},z)\delta T_{b}(\vec{k},z)} \right) \right\rangle\, ,
\end{equation}
where $\Re(x)$ denotes the real part of $x$. As with the 21cm auto-spectrum, it is more convenient to work with the normalized LBG-21cm cross-spectrum
\begin{equation}
P^{\rm norm}_{\rm LBG-21cm}(k,z)=\frac{1}{\overline{\delta T_b}(z)}\left\langle \Re \left( {\delta_{\rm LBG}^*(\vec{k},z)\delta T_{b}(\vec{k},z)} \right) \right\rangle\, .
\end{equation}
The average is performed over all modes, $\vec{k}$, which fall within the spherical shell $[k-\triangle k/2,k+\triangle k/2]$. We use the value of the fundamental mode, $k_{\rm F}=2\pi/L$ (with $L$ being the size of the simulation box), as the width of the interval $\triangle k$.

\begin{figure}
\begin{center}
\includegraphics[width=1.0\textwidth]{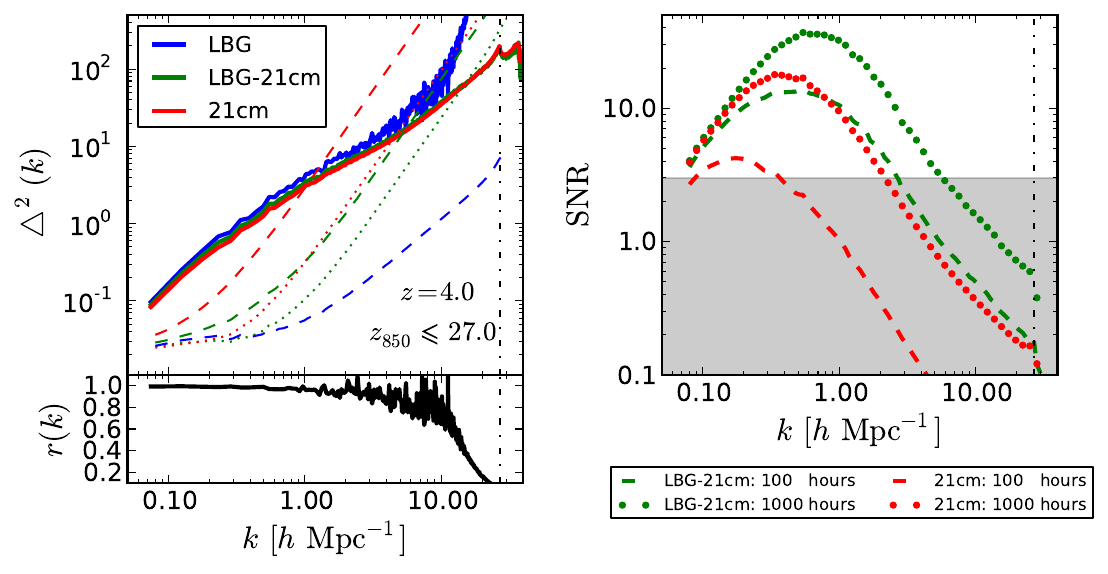}\\
\end{center}
\caption{\textit{Left:} Dimensionless power spectra of LBGs (solid blue) and 21cm intensity (solid red), together with the LBG-21cm cross-spectrum (solid green), at $z=4$. The errors in each quantity are displayed for observing times with the SKA1-LOW array of 100 hours (dashed lines)  and 1000 hours (dotted lines). An HOD model was used to populate the dark matter halos of the N-body simulation with LBGs of magnitude $z_{850} \leqslant 27$, and HI was assigned to the CDM particles belonging to the halos using the Bagla et al. (2010) \cite{Bagla_2010} model. The cross-correlation coefficient is shown in the bottom panel. All the quantities are computed in redshift-space. \textit{Right:} Signal-to-noise ratio (SNR), defined as the ratio between the power spectrum and its error, for 100 hours (dashed lines) and 1000 hours (dotted lines) of observation time. The shaded region shows the places where the SNR is less than 3. In all panels, the dot-dashed vertical line shows the wavenumber of the Nyquist frequency.}
\label{cross_correlation}
\end{figure}

The upper-left panel of Fig. \ref{cross_correlation} compares the various (dimensionless) auto- and cross-spectra at $z=4$, including the 21cm auto-spectrum, $\triangle^2_{\rm 21cm}(k)=k^3P^{\rm norm}_{\rm 21cm}(k)/2\pi^2$ (solid red line); the LBG power spectrum, $\triangle^2_{\rm LBG}(k)=k^3P_{\rm LBG}(k)/2\pi^2$, for a mock catalogue of LBGs with magnitudes $z_{850}\le27.0$ (solid blue line); and the LBG-21cm cross-spectrum, $\triangle^2_{\rm LBG-21cm}(k)=k^3P^{\rm norm}_{\rm LBG-21cm}(k)/2\pi^2$ (solid green line). Note that we have subtracted the shot noise level for the LBG auto-spectrum and that, for simplicity, instrumental effects have not been added to the simulated 21cm intensity maps. Somewhat surprisingly, at this particular redshift, and for this particular LBG population, all the power spectra are very similar in terms of shape and amplitude.

The bottom-left panel of Fig. \ref{cross_correlation} shows the cross-correlation coefficient, defined as
\be
r(k)=\frac{P_{\rm LBG-21cm}(k)}{\sqrt{P_{\rm 21cm}(k)P_{\rm LBG}(k)}}~.
\e
Both fields are strongly correlated, with a value of $r$ close to unity on almost all scales. The sudden drop in the cross-correlation coefficient for $k\gtrsim15~h\,{\rm Mpc}^{-1}$ is due to the increase in power in the LBG power spectrum, which is not fully trustworthy on those scales as they are very close to the Nyquist frequency (vertical dot-dashed line), and are completely dominated by the shot-noise level of the LBGs.

The spherically-averaged power spectrum, $P_A(k)$, (where $A$ stands for LBG, 21cm, or LBG-21cm) has error
\be
\frac{1}{\sigma^2_A(k)}=\frac{2\pi k^2dkV_{\rm survey}}{(2\pi)^3}\int_0^{\pi/2}\frac{\sin(\theta)d\theta}{\sigma_A^2(k,\theta)},
\label{total_error}
\e
where $V_{\rm survey}=D^2 \triangle D (\lambda^2/A)$ is the survey volume, with $D$ representing the comoving distance to the redshift of observation, and $\triangle D$ the comoving distance associated with the bandwidth, $B$. $\lambda$ is the wavelength of observation, and $A$ is the collecting area of a single antenna/station. In the Gaussian limit (see Appendix \ref{sec:appendix} for a detailed discussion of the validity of the Gaussian limit), the errors on the 21cm, LBG, and LBG-21cm power spectra are given by \cite{Furlanetto_2007,Lidz_2009,Smith_2009}
\begin{eqnarray}
\label{error1}
\sigma^2_{\rm LBG}(k,\theta)&=&\left[P_{\rm LBG}(k,\theta)+n_{\rm LBG}^{-1}\right]^2\\
\label{error2}
\sigma^2_{\rm 21cm}(k,\theta)&=&\left[P_{\rm 21cm}(k,\theta)+\frac{T_{\rm sys}^2}{2Bt_0}\frac{D^2\triangle D}{n(k_\bot)}\left(\frac{\lambda^2}{A_e}\right)^2\right]^2\\
\sigma^2_{\rm LBG-21cm}(k,\theta)&=&\frac{1}{2}\left[P_{\rm LBG-21cm}^2(k,\theta)+\sigma_{\rm LBG}(k,\theta)\sigma_{\rm 21cm}(k,\theta)\right]~,
\label{error3}
\end{eqnarray}
where $n_{\rm LBG}$ is the number density of LBGs and $t_0$ is the total observation time. $T_{\rm sys}$ is the system temperature of the radio telescope, which can be written as $T_{\rm sys}=T_{\rm rcvr}+T_{\rm sky}$, with $T_{\rm sky}\approx 60(300{\rm MHz}/\nu)^{2.55}$ K. For SKA1-LOW, $T_{\rm rcvr}=0.1T_{\rm sky}+T_{\rm inst}$, with $T_{\rm inst}=40$ K, whereas for SKA1-MID, $T_{\rm rcvr}=28$ K. The number density of interferometer baselines sensitive to the transverse mode $k_\bot$ is given by $n(k_\bot)$, which depends on the spatial distribution of the antennae; the baseline density distributions for SKA1-LOW and MID are shown in Appendix \ref{sec:SKA1-low}. $A_e$ represents the effective collective area of a station, with $A_e=140~{\rm m}^2$ for SKA1-MID whereas for SKA1-LOW it can be expressed as,
\be 
A_e(\nu) = A_{e, {\rm crit}}\times\left\{ 
  \begin{array}{l l}	
    (\nu_{\rm crit}/\nu)^2 ~~~~& \nu\textgreater\nu_{\rm crit}\\
    \\
    1 & \nu\leqslant\nu_{\rm crit},\\
  \end{array} \right.
\e
where $\nu_{\rm crit}=110$ MHz and $A_{e,{\rm crit}}=925~{\rm m}^2$. We consider two total observation times, 100 hours and 1000 hours, a bandwidth of 32 MHz and take a width of the k-interval equal to $dk=k/5$. In what follows, we focus on SKA1-LOW as it will probe the full range of redshifts that we are interested in here, although we also compute the errors for SKA1-MID at $z=3$ since its band extends out to there. The specifications of the LOW and MID arrays are summarized in Table \ref{radio_telescope_tab}.

The errors on the LBG and LBG-21cm power spectra are computed assuming a LBG survey volume equal to the one of the 21cm field, i.e. $V_{\rm LBG}=D^2\triangle D(\lambda^2/A)$. This implies a field of view of 2.7 deg$^2$ ($z=3$), 4.2 deg$^2$ ($z=4$) and 6.0 deg$^2$ ($z=5$) for SKA1-LOW and 6.9 deg$^2$ ($z=3$) for SKA1-MID (see Table \ref{radio_telescope_tab}), with a depth of 32 MHz. The LBG number density, needed to compute the shot-noise contribution to the overall error, is taken from the last column in Table \ref{HOD_table}.

For simplicity we have neglected the angular dependence of the power spectra, i.e. we have assumed that $P_{\rm LBG}(k,\theta)=P_{\rm LBG}(k)$, $P_{\rm 21cm}(k,\theta)=P_{\rm 21cm}(k)$ and $P_{\rm LBG-21cm}(k,\theta)=P_{\rm LBG-21cm}(k)$. We explicitly tested this assumption by computing the noise in the power spectra using the full angular dependence that we obtain from our N-body simulation. Our results confirm that neglecting the angular dependence of the power spectra is a very precise approximation. We have also assumed that the LBGs are spectroscopically detected; that is, with negligible errors on their measured redshifts. Finally, we have also considered that the survey volume is the same for both the 21cm and LBGs.

\begin{figure}[t]
\begin{center}
\includegraphics[width=1.0\textwidth]{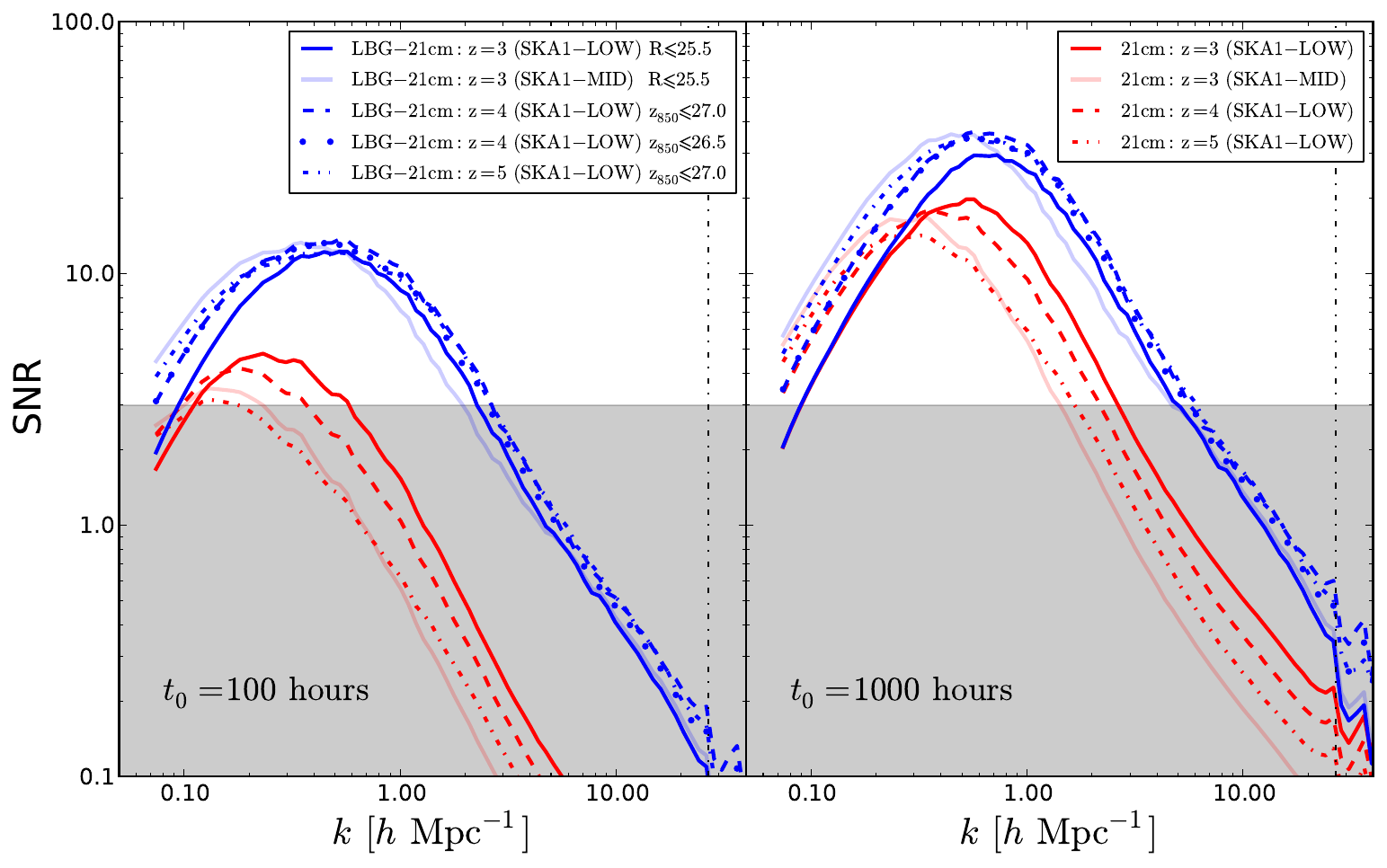}\\
\end{center}
\caption{Signal-to-noise ratio (SNR) of the 21cm auto-power spectrum (red lines) and the LBG-21cm cross-power spectrum (blue lines) for 100 hours (left panel) and 1000 hours (right panel) of observations. The light colored lines show the results at $z=3$ for SKA1-MID while the dark colored lines display the results at $z=3-5$ for SKA1-LOW. The redshift and LBG population considered for the SNR of the auto- and cross-power spectrum is shown in the legend. The shaded region denotes where the SNR is less than 3. The dot-dashed vertical lines show the wavenumber of the Nyquist frequency.}
\label{cross_correlation_SN}
\end{figure}

In the upper-left panel of Fig. \ref{cross_correlation}, the errors on the measured power spectra are shown for total observation times of both 100 and 1000 hours with SKA1-LOW. (Clearly, the error on the LBG auto-spectrum does not depend on the total observing time of the radio telescope.) On small scales, the error on the 21cm auto-spectrum is dominated by the system noise, which scales with the observing time as $1/t_0$, whereas for the cross-spectrum goes as $1/\sqrt{t_0}$. As such, the error on the 21cm auto-spectrum is reduced by a factor of 10 by increasing the observing time from 100 hours to 1000 hours, but the cross-spectrum error will only improve by a factor of $\sim3$.

In the right panel of Fig. \ref{cross_correlation} we show the signal-to-noise ratio (SNR) of the 21cm power spectrum (red lines) and the LBG-21cm cross-power spectrum (green lines) for both 100 hour (dashed lines) and 1000 hour (dotted lines) observations with SKA1-LOW. The 21cm auto-spectrum will barely be detected with a SNR higher than 3 for 100 hours of observation, while for 1000 hours it can be detected down to $k\sim2~h{\rm Mpc}^{-1}$. Conversely, the LBG-21cm cross-spectrum can be detected up to $k\sim2~h{\rm Mpc}^{-1}$ in only 100 hours of observation, with values of $k$ as high as $k\sim6~h{\rm Mpc}^{-1}$ being reachable in 1000 hours. On large scales, and for large observing times, the error budget is dominated by sample variance, and thus the SNR of both the auto- and cross-spectra depends only on the survey volume.

For both observing times considered here, we find that the SNR for the cross-spectrum will be higher than the 21cm auto-spectrum for all relevant wavenumbers. The benefits of targeting the cross-spectrum are therefore twofold: the 21cm signal can be detected with a higher SNR, and down to much smaller scales ($\sim\!10$ times smaller for 100 hours and $\sim\!3$ times for 1000 hours).

We now study the dependence of the LBG-21cm cross-spectrum sensitivity on the LBG population by creating a mock LBG catalogue at $z=4$ containing galaxies with magnitudes $z_{850}\leqslant26.5$ and repeating the above analysis. The SNR for the LBG-21cm signal is shown in Fig. \ref{cross_correlation_SN} (dotted blue line) for both 100 hours (left panel) and 1000 hours (right panel) of observations. The SNR of the LBG-21cm cross-power spectrum for LBG galaxies with $z_{850}\leqslant27$ at $z=4$ (dashed blue line) is shown for comparison. The SNR is not changed much by using brighter LBGs at fixed redshift, the reason being that while the amplitude of the LBG-21cm cross-spectrum increases for brighter galaxies, as they are more strongly clustered, the error also increases as the LBG number density decreases for brighter galaxies.

Next, we investigate the dependence of the cross-spectrum sensitivity on redshift by creating mock LBG catalogues at $z=3$ and $z=5$, containing galaxies with magnitudes $R\le25.5$ and $z_{850}\leqslant27$, respectively. Fig. \ref{cross_correlation_SN} shows the SNR of the 21cm auto-spectrum, for SKA1-LOW, at $z=3$ (solid red line), $z=4$ (dashed red line) and $z=5$ (dot-dashed red line). Results for 100 hours of total observing time are displayed in the left panel, and for 1000 hours in the right. As expected, the SNR of the 21cm power spectrum decreases with increasing redshift, as the system temperature increases. The SNR of the LBG-21cm cross-spectrum is also shown in Fig. \ref{cross_correlation_SN}, for the catalogue at $z=3$ (solid blue line) and $z=5$ (dot-dashed blue line). On small scales, the cross-spectrum SNR does not vary significantly with redshift. This can be understood if we take into account the two effects discussed in the previous paragraph; while the LBG number density decreases at higher redshift, so that the shot noise affecting the LBG power spectrum increases, the amplitude of the LBG-21cm cross-spectrum also increases as galaxies of the same magnitude are more strongly clustered at higher redshift. We note that, on large scales, the SNR of the auto- and cross-power spectrum increases with redshift because the survey volume grows with redshift (mainly due to the increase in the field of view).

Finally, we analyze the dependence of the SNR, for both the 21cm auto-spectrum and LBG-21cm cross-spectrum at $z=3$, on the radio telescope used. In Fig. \ref{cross_correlation_SN} we also show the errors on the auto- (solid light red lines) and cross-spectrum (solid light blue lines) using the specifications of SKA1-MID for the LBG catalogue at $z=3$. For the 21cm auto-spectrum we find that the SNR will be slightly lower than that achieved with SKA1-LOW, while the cross-spectrum SNR on small scales will be very similar for both arrays. Note that the larger survey volume available with MID will allow both the auto- and cross power spectra to be measured, on larger scales, with higher precision, however.

\section{Dependence on the HI model}
\label{sec:HI_modeling}

We now study the dependence of our results on the choice of model for the spatial distribution of neutral hydrogen. So far we have assigned HI to dark matter halos using the Bagla et al. \cite{Bagla_2010} method. In \cite{Villaescusa_2014a}, high-resolution hydrodynamic N-body simulations were used to show that while this method is able to reproduce the DLA column density distribution function, the predicted value of the DLA bias is in strong tension with recent measurements by the BOSS collaboration \cite{Font_2012}. A different method, dubbed \textit{halo-based model 2} in \cite{Villaescusa_2014a}, and inspired by the analytic model of \cite{Barnes_2014}, was able to reproduce both the bias and the column density distribution function of the DLAs, however. In this section we investigate how our results change when the latter method is used to model the distribution of HI.

We begin by briefly describing the halo-based model 2 (see \cite{Villaescusa_2014a} for further details). In this method, as with the Bagla method, we assume that all HI resides within dark matter halos. Now, though, the HI mass within a given dark matter halo, $M_{\rm HI}(M)$, is assumed to only depend on its mass, and is modeled as
\begin{equation}
M_{\rm HI}(M)=\alpha f_{\rm H,c}\exp{\left[ -\left(\frac{v_c^0}{v_c}\right)^\beta\right]}M~,
\label{M_HI_Paco}
\end{equation}
where $f_{\rm H,c}=0.76\Omega_{\rm b}/\Omega_{\rm m}$ is the cosmic
hydrogen mass fraction, $v_c$ is the halo circular velocity, and
$\alpha$, $\beta$, and $v_c^0$ are the free parameters of the model. Their values are chosen to reproduce the column density distribution function and bias of the DLAs; at $z=4$ their values are: $\alpha=0.34$, $\beta=3$, and $v_c^0=37~{\rm km/s}$. In \cite{Villaescusa_2014a}, the authors phenomenologically tuned the HI density profile within the dark matter halos to reproduce the DLA column density distribution, but here -- constrained by the fact that the simulations carried out for this work are not hydrodynamical -- we follow the spirit of the Bagla method and distribute all of the HI mass of a given dark matter halo equally amongst its constituent particles.

\begin{figure}
\begin{center}
\includegraphics[width=1.0\textwidth]{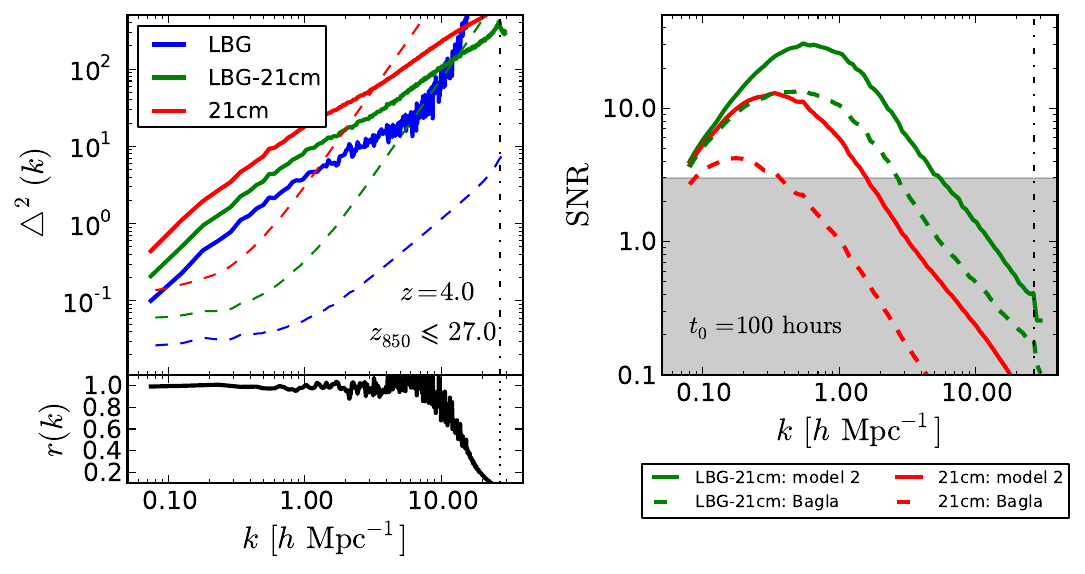}\\
\end{center}
\caption{Same as Fig. \ref{cross_correlation}, but now using the halo-based model 2 for the spatial distribution of HI (see text for details). The observing time is 100 hours. In the right panel we show the SNR of the 21cm power spectrum (red lines) and LBG-21cm cross-power spectrum (green lines) for both the halo-based model 2 (solid lines) and the Bagla model (dashed lines).}
\label{cross_correlation_Paco}
\end{figure}

Using this method to assign HI to the simulated dark matter halos, we cross-correlated the 21cm field with the distribution of LBGs with magnitudes $z_{850}\le27$, all at $z=4$. The results are shown in Fig. \ref{cross_correlation_Paco}, with errors again computed for SKA1-LOW. As expected, we find that the amplitude of the 21cm power spectrum predicted by this method is much higher than for the Bagla method, by a factor of about 6. This happens because the halo-based method 2 assigns a much higher HI mass to the most massive halos which boost the amplitude of the 21cm power spectrum. Since the faint LBGs we consider here tend to live in low mass halos while the halo-based model 2 places the HI mainly in the most massive halos, it is not surprising that the amplitude of the 21cm power spectrum is higher than the one of the LBGs. The cross-correlation coefficient does not change, however, and the 21cm and LBG fields remain very strongly correlated on large scales. The correlation does weaken on small scales though, as a consequence of the increase in the LBG auto-power for wavenumbers where the LBG shot noise begins to rise.

The boost in the amplitude of the 21cm auto-spectrum for halo-based model 2 is translated into a higher SNR on small scales, as the uncertainty is dominated by the system noise there. Conversely, the error on the large-scale LBG-21cm cross-spectrum is dominated by sample variance, and so the SNR does not change much despite the increased cross-power amplitude. For the small-scale cross-spectrum, however, the SNR does not increase as much as for the 21cm auto-spectrum; the error budget is dominated by the system noise of the 21cm observations but, for strongly correlated fields, the amplitude of the cross-spectrum only increases with the square root of the 21cm power spectrum, $P_{\rm LBG-21cm}(k)\simeq\sqrt{P_{\rm 21cm}(k)P_{\rm LBG}(k)}$. As such, on small scales, the SNR of the 21cm power spectrum increases linearly with the amplitude of the 21cm power spectrum, while for the cross-spectrum it only increases as the square root.

\section{Effect of foreground cleaning} 
\label{sec:fg}

A particularly challenging problem for 21cm intensity mapping is the presence of radio foregrounds,
both galactic and extragalactic, with amplitudes several orders of magnitude larger than the cosmological HI
signal. The properties of the various foreground signals have been described in detail in the literature
\cite{2002ApJ...564..576D,2005ApJ...625..575S,2009A&A...500..965B,2013ApJ...769..154M}, and several
ways of removing them have been proposed, particularly for the EoR
regime \cite{2011PhRvD..83j3006L,Masui_2013,2014MNRAS.441.3271W,2014ApJ...781...57S}. Most removal methods
rely on the assumed smoothness of the foreground frequency spectra to disentangle them from the cosmological
signal and, as long as this holds, the foreground cleaning should be successful in a wide range of scales.
Due to a lack of appropriate large-area multi-frequency observations of the radio sky, however, there still
exist important uncertainties regarding the structure of these foregrounds, and the effects of foreground
cleaning on the recovered power spectrum. Since these issues could have a significant effect on the recovered
auto- and cross-spectra, we have attempted to quantify the impact of foreground cleaning on our results.

We first generated a realisation of a single foreground source as a function of frequency and angle in the
volume spanned by the simulation box at $z=3$. For this, we used a flat-sky version of the public code from
\cite{2014MNRAS.444.3183A} to simulate a foreground source with the amplitude, angular structure, and spectral
index of unpolarized galactic synchrotron (i.e. with the parameters $A$, $\beta$ and $\alpha$ taken from the
first row of Table 1 in \cite{2014MNRAS.444.3183A}),
and with the frequency correlation length of point sources (i.e. with
$\xi$ taken from the second row of the same table). While this set of parameters does not correspond
to a particular physical foreground source, it combines the amplitude and spatial structure of galactic
synchrotron (by far the largest foreground) with the degree of frequency decorrelation of point
sources (the least correlated). The results of subtracting this foreground should
therefore give a conservative estimate of the overall effect in a more realistic scenario.

We produced `observed' maps by adding the foreground realisation to the 21cm signal, and then tried to recover
the cosmological signal by applying two of the blind
foreground cleaning methods studied in \cite{2015MNRAS.447..400A} -- polynomial fitting and Principal Component
Analysis (PCA). We refer the reader to this paper for detailed descriptions of the methods. To avoid edge
effects close to the boundaries of the frequency band \cite{2015MNRAS.447..400A}, we applied the cleaning
methods to an extended data cube generated by doubling the size of the simulation box in the radial (line of
sight) direction. This was done by replicating the simulation of the cosmological signal (both the HI temperature
and galaxy positions) in this direction using the periodic boundary conditions of the N-body simulation, and then
directly generating the foreground realisation in this extended frequency range. In total 512 frequency bins were
used to perform the foreground cleaning. Note that while the cleaning methods were applied on the extended range,
the subsequent analysis of the power spectrum was performed only in the original simulation volume.

\begin{figure}[t]
  \begin{center}
    \includegraphics[width=1.0\textwidth]{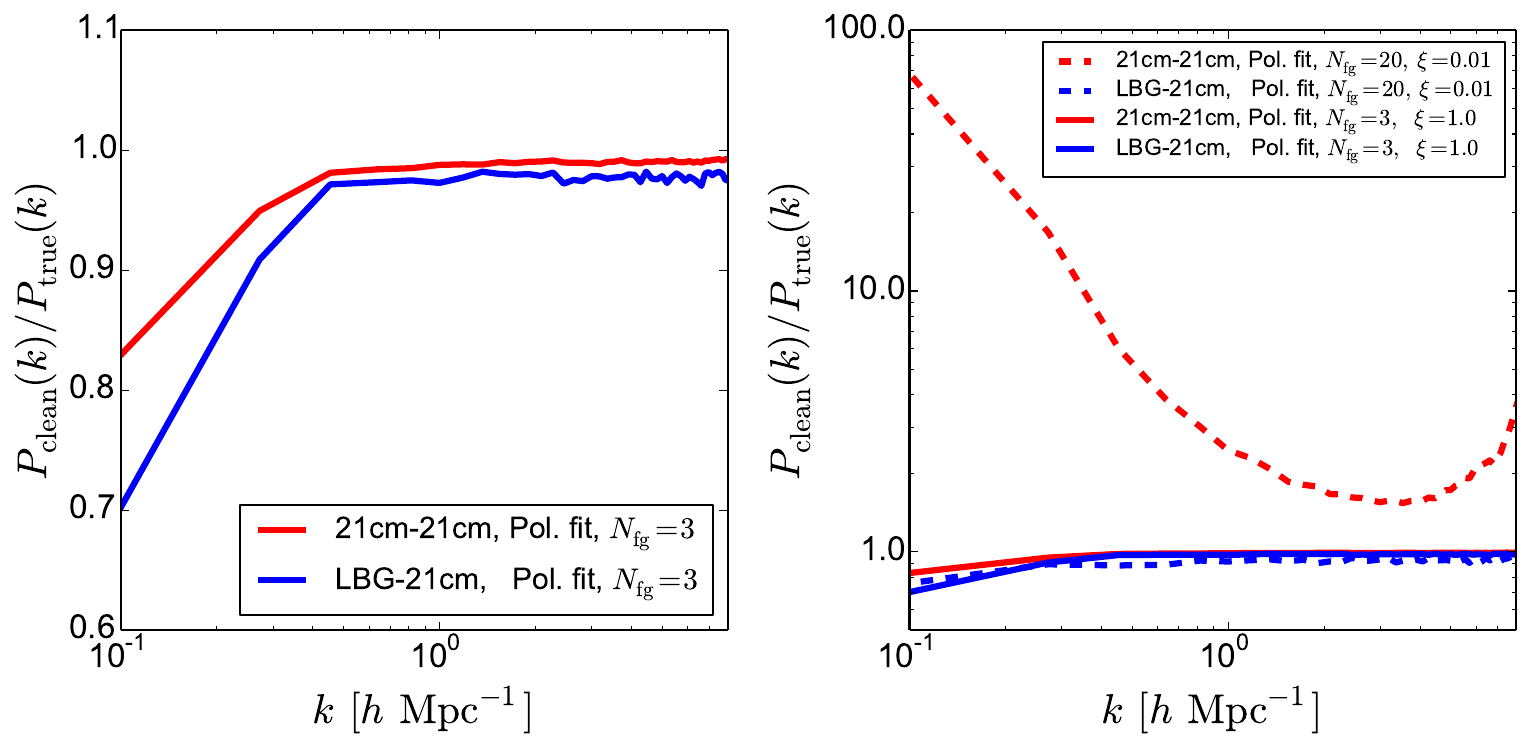}
  \end{center}
  \caption{{\sl Left panel:} Ratio of the $z=3$ power spectrum of the foreground-cleaned data to the true power
  spectrum for the 21cm autocorrelation (red line) and LBG-21cm cross-correlation (blue line). These results were
  found for a smooth foreground source with $\xi=1.0$ using a polynomial fitting technique, fitting 3 degrees of
  freedom for every line of sight. {\sl Right panel:} The same quantities, but now with an additional non-smooth
  foreground with $\xi=0.01$. The red and blue solid lines show the result for the auto- and cross-correlations
  respectively, while the dashed lines show the results for the smooth foreground only.}
  \label{fig:fgrm}
\end{figure}

For each method, we determined the optimal number of foreground degrees of freedom to subtract, $N_{\rm fg}$, by
finding the value of $N_{\rm fg}$ that minimises the induced bias on the estimated power spectrum. Both methods
were able to clean the foregrounds to the same level with an optimal value of $N_{\rm fg}=3$. The left panel in
Fig. \ref{fig:fgrm} shows the ratio of the power spectrum of the foreground-cleaned maps to the true (i.e.
foreground-free) power spectrum for both the 21cm auto-correlation and the cross-correlation with LBGs. The
effect of foreground removal is minimal on small scales, but
becomes significant for larger ones; this is because long-wavelength cosmological modes vary relatively smoothly
in frequency-space, and so are partially subtracted along with the smooth foreground signal.

It is interesting to note that, due to the partial subtraction of the 21cm signal on large radial scales, the
estimated cross-spectrum is biased in the same way as the auto-spectrum.
A possible way of avoiding this bias would be to cross-correlate the LBG positions with the uncleaned intensity
maps (or maps cleaned to a sub-optimal level). Since the LBGs
and the foregrounds are uncorrelated, this estimator would be completely unbiased. Due to the large amplitude of
the foregrounds, however, the variance for this estimator would be completely dominated by the variance of the
foregrounds, rendering it useless for making precision measurements.
It is also worth noting that although the bias induced by foreground cleaning on the recovered power spectra is
similar for both the auto- and cross- correlations, the effect on the cross-power spectrum is slightly larger
on all scales.
Although a priori this result might seem contradictory, since any signal loss will be squared in the auto-correlation,
it is actually reasonable: the cleaned maps are a combination of both the cosmological signal and foregrounds (with
hopefully a very small contribution from the latter). Thus,
the auto-correlation receives positive contributions from both components, while the cross-correlation is unaffected
by the residual foregrounds. Hence the bias in the cross-correlation is a more faithful measure of the amount of
signal loss after foreground cleaning.

Nevertheless, compared to the auto-correlation, the cross-correlation of the 21cm signal with any other cosmological
tracer, such as LBGs, should be extremely robust against non-smooth foregrounds or instrumental effects \cite{Masui_2013}.
An important concern for intensity mapping experiments, for example, is the leakage of polarized synchrotron foregrounds
into the total intensity data. Faraday rotation renders the frequency structure of polarized synchrotron emission highly
non-trivial, and because of our lack of understanding of the galactic magnetic field causing this effect, the
uncertainties on the degree of frequency decorrelation of this foreground are quite large \cite{2013ApJ...769..154M}.
In order to show the usefulness of cross-correlating the HI signal with other tracers as a way of circumventing this kind
of problem, we generated another realisation of a foreground source, corresponding to a foreground with an amplitude
similar to that of the HI signal, with a small frequency correlation length of $\xi=0.01$. After adding this to the
observed maps, we attempted to subtract it by applying a polynomial fitting algorithm again. The results are shown in the
right panel of Fig. \ref{fig:fgrm}; while it is impossible to recover the correct 21cm auto-spectrum in the presence of
this foreground -- regardless of the number of parameters used to fit each line of sight -- only a mild level of
foreground subtraction is required in order to recover the cross-correlation to an acceptable level.

It is worth noting that in this analysis we have assumed no correlation between foregrounds and cosmological signal.
Strictly speaking this is not the case: LBGs show detectable radio emission \cite{2008ApJ...689..883C}, and in general
extragalactic foregrounds (mainly radio point sources) probe the same large-scale structure as any cosmological signal.
The impact of these correlations was studied by \cite{2014MNRAS.444.3183A} in the context of the 21cm auto-correlation,
showing that the extra effects of a foreground source that is correlated with the cosmological signal are negligible
compared with the foreground residuals already present for an uncorrelated source. Furthermore, the main contribution
to these foregrounds comes from low-redshift galaxies, and therefore the overlap with the measurements at $z\gtrsim3$
proposed here should be negligible.

The results shown in this section should be understood as giving a qualitative picture of what the actual effect would be
in a realistic scenario. The performance of any foreground subtraction technique depends not only on the precise
statistical properties of the foregrounds, but also on instrumental parameters such as the angular and frequency resolution,
the observed fraction of the sky, the level of instrumental noise, and so on, which we have not included in our simulation.
A quantitative estimation of these effects lies beyond the scope of this paper, and is left for future work.

\section{Discussion and conclusions}
\label{sec:conclusions}

Precision cosmological constraints are currently provided by a variety of observables, including the cosmic microwave background (CMB), supernova distance measurements, and the clustering of galaxies (including baryon acoustic oscillations, BAO). A new technique, 21cm intensity mapping (IM), is expected to contribute significant improvements to these constraints in a number of ways (see for instance \cite{Bull_2014}).

Intensity mapping will allow us to probe the matter distribution on redshifts not probed by any other cosmological observable, in particular, will let us to constrain the matter power spectrum on the post-reionization window $2\lesssim z \lesssim 6$. Unfortunately, intensity maps are strongly contaminated by the emission from galactic and extra-galactic foregrounds, which is several orders of magnitude stronger than the cosmological 21cm signal. A way of mitigating foreground contamination is to cross-correlate the 21cm intensity maps with galaxy catalogues. The foreground signals should mostly be uncorrelated with the cosmological HI and galaxy distributions, and so their effect on the galaxy-21cm cross-correlation is much reduced. The cross-spectrum can therefore be used as a robust way to confirm the detection of the 21cm auto-spectrum, and to otherwise study the signal in the presence of smaller systematic effects. Therefore, the cross-power spectrum is a more suitable quantity to study the IM signal than the 21cm auto-correlation, since it is less affected by the different contaminants of the IM signal.

On the other hand, the LBG-21cm cross-power spectrum can also be used to extract very important cosmological and astrophysical information such as the 21cm bias: on large scales, the cross-power spectrum is expected to follow the shape of the underlying matter distribution, modulo a linear bias factor: $P_{\rm LBG-21cm}(k)=b_x^2P_{\rm m}(k)$, with $P_{\rm m}(k)$ being the matter power spectrum and $b^2_x=b_{\rm LBG}b_{\rm 21cm}r$, where $r$ is the value of the cross-correlation coefficient on large scales. Thus, knowing the bias of the LBGs, and assuming that on large scales both fields will be completely correlated (i.e. $r=1$, see for instance Fig. \ref{cross_correlation}), it is then possible to estimate the bias of the 21cm, and compare it with the value obtained from 21cm auto-correlation measurements. The value of $b_{\rm 21cm}(z)$ encodes important astrophysical information on the way galaxies accrete gas over cosmic time, and how this gas is converted into stars. Thus, $b_{\rm 21cm}$ is a fundamental quantity targeted by theories trying to explain galaxy formation and evolution.

In this paper we investigated the cross-correlation between 21cm intensity fluctuations and Lyman Break Galaxies (LBGs) at $z \sim 3 - 5$, including the detectability of the cross-spectrum with the future SKA1-LOW and SKA1-MID instruments, and the effects of foreground cleaning on both the 21cm auto-spectrum and the LBG-21cm cross-spectrum. We also studied the dependence of the signal-to-noise ratio (SNR) of the LBG-21cm cross-spectrum on the galaxy population, redshift, experimental set-up and the method used to model the spatial distribution of neutral hydrogen. The 21cm intensity field was generated by assigning HI, a posteriori, to the CDM particles belonging to dark matter halos of an N-body simulation, using the Bagla et al. model \cite{Bagla_2010}, and the spatial distribution of LBGs was obtained by populating the dark matter halos with galaxies using a halo occupation distribution model (HOD).

We find that the LBG-21cm cross-spectrum can be detected with a much higher SNR (up to 10 times larger) than the 21cm auto-spectrum. For 100 hours of observations with SKA1-LOW, our results show that, depending on redshift, the 21cm auto-spectrum can only be detected with a SNR higher than 3 between $0.1\lesssim k~(h{\rm Mpc}^{-1})\lesssim 0.2-0.4$, whereas the LBG-21cm cross-spectrum will be detected at this level between $0.1 \lesssim k~(h{\rm Mpc}^{-1}) \lesssim 3$. For 1000 hours of observations the 21cm auto-spectrum will only be detected between $0.1\lesssim k~(h{\rm Mpc}^{-1}) \lesssim 2$, while with the cross-spectrum, SKA1-LOW will go down to $k\sim5~h{\rm Mpc}^{-1}$.

We also compared the SNR on the 21cm auto-spectrum and the LBG-21cm cross-spectrum that can be achieved at $z=3$ with SKA1-LOW and SKA1-MID. On small scales, SKA1-LOW will detect the auto-spectrum with higher SNR, while the cross-spectrum will be detected to a similar level by both arrays. On large scales, however, the larger survey volume achievable by SKA1-MID will yield a higher SNR than SKA1-LOW for both the auto- and cross-spectra.

In our calculations, we have assumed that the LBG redshifts are spectroscopically-confirmed. For photometric redshifts, the LBG shot noise errors will increase substantially, significantly degrading the SNR of the cross-spectrum. We also assumed a LBG survey volume equal to that of the 21cm survey\footnote{Notice that the error on the cross-power spectrum scales inversely with volume sampled by the intersection of both the 21cm and the LBG surveys.}, which for SKA1-LOW corresponds to an area of $\sim3-6~\deg^2$ at redshifts $z=3-5$ (see table \ref{radio_telescope_tab}). We note that current LBG surveys cover areas comparable to those: $\sim9~{\rm deg^2}$ with photometric redshifts \cite{Bian_2013} and $\sim4~{\rm deg^2}$ with spectroscopic redshifts \cite{Bielby_2013}. Moreover, in the near future, the Hyper Suprime-Cam\footnote{http://www.naoj.org/Projects/HSC/} and the Prime Focus Spectrograph\footnote{http://sumire.ipmu.jp/en/2652} on the Subaru telescope will detect LBGs in the redshift range $2~\textless~z~\textless~5$ in a very wide area ($\sim30~{\rm deg^2}$). 

For the redshifts and LBG populations explored in this paper, we find that the SNR of the LBG-21cm cross-spectrum depends only weakly on redshift and on the limiting magnitude of the LBGs. This is due to two competing effects. On the one hand, the errors on the 21cm signal due to instrumental noise increase with redshift, as does the LBG shot noise (for galaxies with the same magnitude). On the other hand, galaxies at higher redshift, or brighter galaxies at the same redshift, are more strongly clustered, and so the amplitude of the LBG-21cm cross-spectrum increases. The two effects have similar magnitudes, essentially canceling any effect on the SNR.

We also investigated the dependence of our results on the model used to simulate the spatial distribution of neutral hydrogen. The dark matter halos of our N-body simulation at $z=4$ were populated with HI using two different methods: the Bagla method \cite{Bagla_2010} and the halo-based model 2 method \cite{Villaescusa_2014a}. The amplitude of the 21cm power spectrum is approximately 6 times higher for the halo-based model 2. Using this model, we find that SKA1-LOW will detect the 21cm auto-spectrum at $z=4$ with a SNR above 3 between $0.1\lesssim k~(h{\rm Mpc}^{-1}) \lesssim 2$ with only 100 hours of observations, while the LBG-21cm cross-spectrum can be detected to significantly smaller scales ($k \sim 5~h{\rm Mpc}^{-1}$) with the same observation time. For the redshifts and galaxy populations studied in this paper, we emphasize that the SNR of the LBG-21cm cross-spectrum is always higher than the one of the 21cm auto-spectrum, independently of the model used to simulate the distribution of HI.

Finally, we studied the effects of foreground cleaning on the recovery of the cosmological signal for both the 21cm auto-spectrum and the LBG-21cm cross-spectrum. For a smooth foreground source with a frequency correlation coefficient of $\xi=1.0$, both spectra can be recovered with only a very small bias on small scales, and a slightly larger bias on large scales. The magnitude of the bias is similar for both spectra. In the presence of non-smooth foregrounds, however, the recovered 21cm auto-spectrum is strongly biased, while the cross-spectrum remains essentially unaffected.

\section*{Acknowledgements}
We thank Emiliano Sefusatti, Cristiano Porciani, and Adam Lidz for useful discussions.
The computations in this paper were run on the Odyssey cluster supported by the FAS Division of Science, Research Computing Group at Harvard University. Partial analysis of the simulations was also carried out on SOM2 and SOM3 at IFIC. FVN and MV are supported by the ERC Starting Grant ``cosmoIGM'' and partially supported by INFN IS PD51 ``INDARK'' and by PRIN INAF 2011 ``A complete view on the first 2 billion years of galaxy formation". DA is supported by ERC grant 259505. KKD thanks the Department of Science \& Technology (DST), India for the research grant SR/FTP/PS-119/2012 under the Fast Track Scheme for Young Scientist. PB is supported by ERC grant StG2010-257080. MGS acknowledges support from the National Research Foundation (NRF, South Africa), the South African Square Kilometre Array Project and FCT under grant PTDC/FIS-AST/2194/2012. We acknowledge partial support from ``Consorzio per la Fisica -- Trieste''.

\appendix
\section{SKA configuration}
\label{sec:SKA1-low}

\begin{table}
\begin{center}
{\renewcommand{\arraystretch}{1.2} \begin{tabular}{|l||c||c|c|c|}
\hline
{\bf Telescope}  & {\bf SKA1-MID} & \multicolumn{3}{c|}{\bf SKA1-LOW} \\ \hline\hline
  Redshift & z=3 & z=3 & z=4 & z=5 \\
\hline
Observing time & \multicolumn{4}{c|}{100 / 1000 hours} \\ \hline
Bandwith ($B$) & \multicolumn{4}{c|}{32 MHz} \\ \hline
Frequency range & 350 -- 1050 MHz & \multicolumn{3}{c|}{50 -- 350 MHz}\\ \hline
Number of stations & 254 &\multicolumn{3}{c|}{911}\\ \hline
Station/dish diameter & 15 m & \multicolumn{3}{c|}{35 m}\\ \hline
Field of View (deg$^2$) & 6.9 & 2.7 & 4.2 & 6.0 \\ \hline
System temp. ($T_{\rm sys}$) & 70 K & 82 K & 115 K & 159 K\\ \hline
Effective area ($A_e$) & 140 m$^2$ & 89 m$^2$ & 139 m$^2$ & 200 m$^2$\\ \hline
\end{tabular} }
\caption{\label{radio_telescope_tab}Specifications for the SKA1-MID and SKA1-LOW arrays.}
\end{center}
\end{table}

\begin{figure}
\begin{center}
\includegraphics[width=0.6\textwidth]{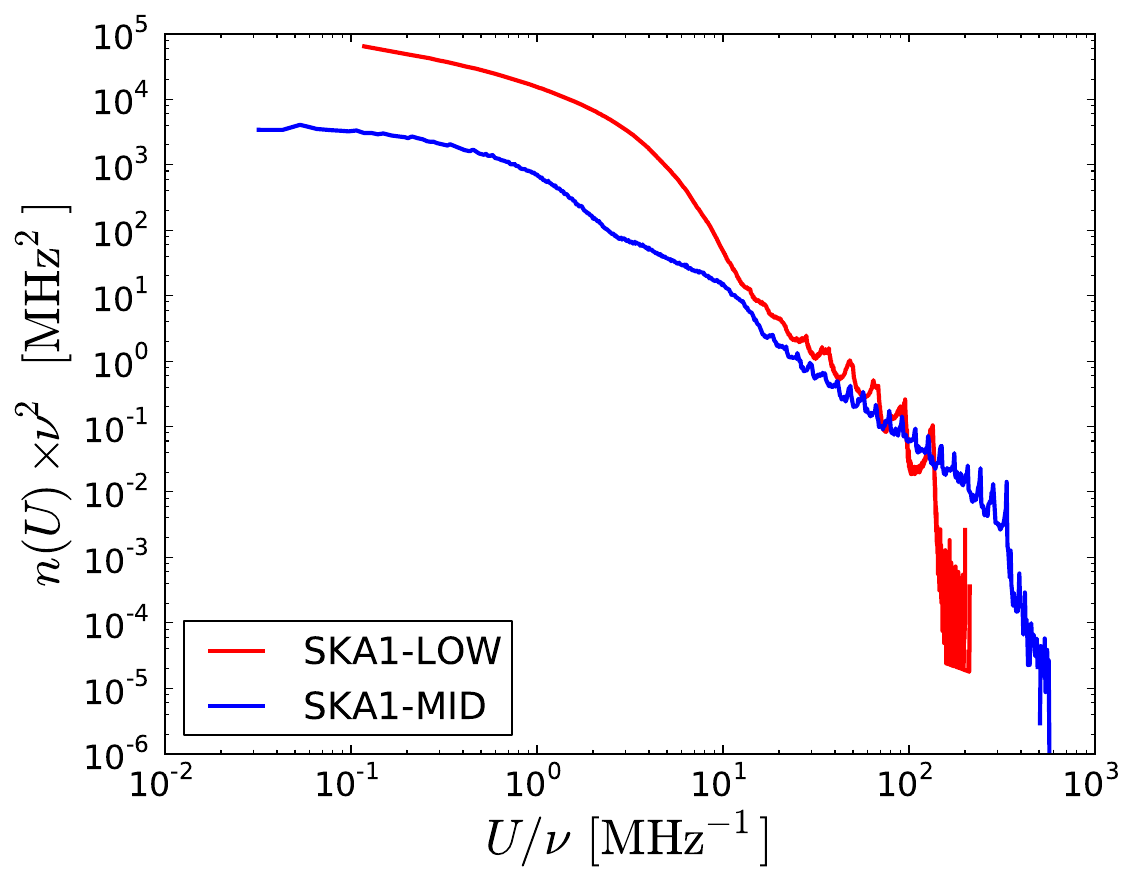}\\
\end{center}
\caption{Baseline density distribution used for SKA1-LOW (red) and SKA1-MID (blue).}
\label{baseline_distribution}
\end{figure}

Table \ref{radio_telescope_tab} displays the assumed specifications for the SKA1-LOW and SKA1-MID arrays. Fig. \ref{baseline_distribution} shows the interferometer baseline density distributions, $n(U)$, used to compute the errors on the 21cm auto-/cross-spectra for the SKA1-LOW and SKA1-MID radio arrays.

\section{Errors on the cross-power spectrum}
\label{sec:appendix}

Eqs. \ref{error1}, \ref{error2}, and \ref{error3} are derived assuming that the fields are Gaussian. 
In this Appendix we investigate whether those expressions are accurate to estimate the error on the LBG-21cm cross-power on the small, fully non-linear, scales that will be probed by the cross-correlation. Here we derive the full non-linear expression for the error on the cross-spectrum and compare it with its Gaussian limit. We demonstrate that on small scales the cross-power spectrum error is dominated by the product of the LBGs power spectrum and the system noise temperature, and that non-linear corrections arising from the cross-power spectrum sample variance are subdominant on those scales.

The LBG power spectrum is defined as $P^d_{\rm LBG}(k)=\langle \delta_{\rm LBG}(\vec{k}) \delta_{\rm LBG}(-\vec{k}) \rangle$, where the superscript $d$ is written to reinforce that it is the power spectrum of a discrete tracer. It can be shown that the relation between the power spectrum of a discrete tracer and the underlying continuous tracer field is given by $P^d_{\rm LBG}(k)=P^c_{\rm LBG}(k)+n_{\rm LBG}^{-1}$ (see for instance \cite{Smith_2009}), where $n^{}_{\rm LBG}$ is the number density of the LBG population.

\begin{figure}[t]
\begin{center}
\includegraphics[width=0.65\textwidth]{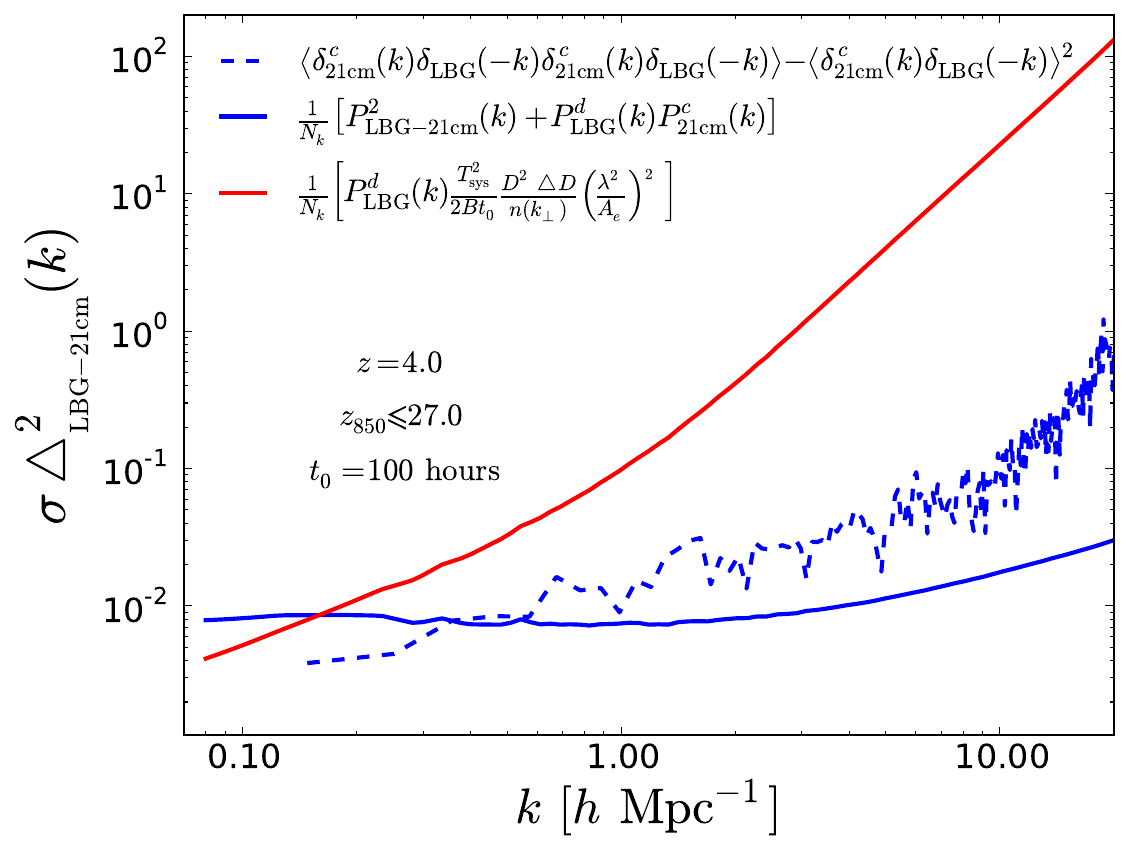}\\
\end{center}
\caption{Contribution of different terms to the total error of the dimensionless LBG-21cm cross-spectrum. The contribution from sample variance is shown with blue lines in the Gaussian limit (solid line) and computed by dividing the simulation box into 8 sub-boxes and measuring the covariance matrix of the cross-power spectrum (dashed line). The solid red line is the contribution arising from the instrumental noise of the array. Calculations are at $z=4$, for LBGs with magnitudes $z_{850}\le27.0$, with the HI modeled using the Bagla et al. method and assuming 100 hours of observations.}
\label{cross_correlation_error}
\end{figure}

Similarly, the 21cm power spectrum is defined as $P^m_{\rm 21cm}(k)=\langle \delta_{\rm 21cm}(\vec{k}) \delta_{\rm 21cm}(-\vec{k})\rangle$, where the superscript $m$ denotes the measured 21cm power spectrum. The observed 21cm fluctuation receives contributions from two sources: $\delta_{\rm 21cm}(\vec{k})=\delta^c_{\rm 21cm}(\vec{k})+\triangle\delta_{\rm 21cm}(\vec{k})$, where $\delta^c_{\rm 21cm}(\vec{k})$ is the contribution from the cosmological 21cm signal, and $\triangle\delta_{\rm 21cm}(\vec{k})$ is the (Gaussian) system noise. For simplicity we neglect errors arising from foreground subtraction. The measured 21cm power spectrum is therefore given by $P^m_{\rm 21cm}(k)=P^c_{\rm 21cm}(k)+\langle \triangle\delta_{\rm 21cm}(\vec{k}) \triangle\delta_{\rm 21cm}(-\vec{k}) \rangle$, with (see \cite{Morales_2005,McQuinn_2006,Geil} for a derivation)
\begin{equation}
\left\langle \triangle\delta_{\rm 21cm}(\vec{k}) \triangle\delta_{\rm 21cm}(-\vec{k}) \right\rangle = \frac{T_{\rm sys}^2}{2Bt_0}\frac{D^2\triangle D}{n(k_\bot)}\left(\frac{\lambda^2}{A_e}\right)^2.
\end{equation}
The various terms are defined in Sec. \ref{sec:results}. The covariance matrix of the cross-spectrum is 
\begin{eqnarray}
\nonumber
C_{\rm LBG-21cm}(\vec{k}_1,\vec{k}_2)&=&\left\langle \delta_{\rm 21cm}(\vec{k}_1)\delta_{\rm LBG}(-\vec{k}_1) \delta_{\rm 21cm}(\vec{k}_2)\delta_{\rm LBG}(-\vec{k}_2) \right\rangle - \\
&&\left\langle \delta_{\rm 21cm}(\vec{k}_1) \delta_{\rm LBG}(-\vec{k}_1)\right\rangle \left\langle \delta_{\rm 21cm}(\vec{k}_2) \delta_{\rm LBG}(-\vec{k}_2)\right\rangle~.
\end{eqnarray} 
Substituting $\delta_{\rm 21cm}(\vec{k})$ by $\delta^c_{\rm 21cm}(\vec{k})+\triangle\delta_{\rm 21cm}(\vec{k})$ in the above expression, one obtains
\begin{eqnarray}
\nonumber
C_{\rm LBG-21cm}(\vec{k}_1,\vec{k}_2)&=&\left\langle \delta^c_{\rm 21cm}(\vec{k}_1) \delta_{\rm LBG}(-\vec{k}_1) \delta^c_{\rm 21cm}(\vec{k}_2)\delta_{\rm LBG}(-\vec{k}_2) \right\rangle - \\
\nonumber
&&\left\langle \delta^c_{\rm 21cm}(\vec{k}_1) \delta_{\rm LBG}(-\vec{k}_1)\right\rangle \left\langle \delta^c_{\rm 21cm}(\vec{k}_2) \delta_{\rm LBG}(-\vec{k}_2)\right\rangle+\\
&&P^d_{\rm LBG}(\vec{k}_2)\frac{T_{\rm sys}^2}{2Bt_0}\frac{D^2\triangle D}{n(k_{1,\bot})}\left(\frac{\lambda^2}{A_e}\right)^2\delta_{\vec{k}_1,-\vec{k}_2}~.
\label{cross_error}
\end{eqnarray} 
The first two terms represent the covariance between the cosmological 21cm signal and the LBG field, and the last term is the instrumental noise. In the Gaussian limit, and for $\vec{k}_1=\vec{k}_2=\vec{k}$, the above expression reduces to
\begin{equation}
C_{\rm LBG-21cm}(k)=\frac{1}{N_k}\left[P^2_{\rm LBG-21cm}(k)+P^d_{\rm LBG}(k)P^c_{\rm 21cm}(k)+
P^d_{\rm LBG}(k)\frac{T_{\rm sys}^2}{2Bt_0}\frac{D^2\triangle D}{n(k_\bot)}\left(\frac{\lambda^2}{A_e}\right)^2\right]~,
\label{cross_Gaussian}
\end{equation} 
which is the equivalent to Eqs. \ref{total_error} and \ref{error3} with $N_k$ being the number of modes. We have quantified the corrections coming from higher order correlations \cite{Smith_2009} to the error on the cross-power spectrum as follows. At $z=4$ we have modeled the spatial distribution of HI using the Bagla et al. model and we used our mock LBG catalogue with galaxies with magnitudes $z_{850}\le27.0$. We then divide the simulation box into 8 sub-boxes and measure the LBG-21cm cross-power spectrum in each of them. Finally, for every wavenumber, we compute the variance of the cross-power spectrum. The results, are shown in Fig. \ref{cross_correlation_error} with a blue dashed line. In that figure we also show with a solid blue line the covariance of the cross-power spectrum in the Gaussian limit, i.e. the first two terms in Eq. \ref{cross_Gaussian}. We find that, on large scales, the errors in the cross-power spectrum arising from the sample variance of the LBG-21cm cross-correlation agree reasonably well between the full covariance matrix and the purely Gaussian part, while on small scales the errors are substantially higher in the when computing the full covariance matrix.

Fig. \ref{cross_correlation_error} also shows with a solid red line the last term of Eq. \ref{cross_Gaussian} (solid red line), due to the instrumental noise (we have assumed 100 hours of observations). On small scales, the contribution of this term is much higher than the one from sample variance, and therefore we conclude that using the Gaussian limit of the expression for the error on the LBG-21cm cross-spectrum is a very good approximation. The reason is simple: on large scales the errors are well-described by Gaussian fields, while on small scales the error budget is dominated by the system noise, which is assumed to be Gaussian.

\bibliographystyle{JHEP}
\bibliography{Bibliography} 

\end{document}